\documentclass[a4paper,11pt]{article}
\usepackage{jinstpub} 
\usepackage{multirow}
\usepackage{wrapfig}
\usepackage{bm}

\title{A space-time tracking algorithm for high occupancy events at future colliders}

\author[a]{Massimo Casarsa,}
\author[b]{Sergo Jindariani}
\author[b,1]{and Luciano Ristori\note{Corresponding author.}}

\affiliation[a]{INFN Sezione di Trieste, via A. Valerio 2, I-34127 Trieste, Italy}
\affiliation[b]{Fermilab, P.O. Box 500, Batavia, Illinois, 60510, USA}

\emailAdd{luciano@fnal.gov}

\abstract{

   We propose to explore the potential advantages of a new class of tracking algorithms loosely inspired by the Hough transform concept
   and where we include the time of arrival of each hit as an additional coordinate to be treated in the same way as a spatial coordinate.   
   A remarkable property of this algorithm is that the execution time is proportional to the total number of hits to be processed, making it particularly attractive for high occupancy situations expected at future colliders.
   The particular structure of the algorithm also lends itself naturally to parallel hardware implementations which, combined to its intrinsic flexibility, should provide a powerful tool for triggering at future colliders. To probe the effectiveness of the algorithm, we apply it to a quasi-realistic simulated environment of a possible future muon collider experiment and report the performance.
   
}

\keywords{Pattern recognition, Data processing methods, Trigger algorithms, Charged track reconstruction}

\begin{document}
\maketitle
\flushbottom

\section{Introduction}
\label{sec:Introduction}
  
    For the future generation of particle colliders, efficient track finding algorithms will be crucial for both online and offline processing of data. All efforts to improve on the quality and performance of tracking algorithms are therefore amply motivated. 
    Track reconstruction involves a series of steps designed to optimize performance while minimizing execution time.
    The first step is the {\em Pattern Recognition} phase, in which we try to identify all the hits that are consistent with the hypothesis of being originated by the same single particle track, and discard most of noise hits.
    In the second step, each group of hits identified during Pattern Recognition, undergoes the {\em Track Fitting} phase, where the most probable values of the track parameters are determined.
    The final step involves using the track parameters, the associated hits, and the fit quality information to eliminate fake and duplicate tracks, ensuring the accuracy and integrity of the reconstruction.  
    This paper is focused on the description of an innovative algorithm that can be used to perform the Pattern Recognition phase and of a first set of tests aimed at demonstrating its effectiveness and desirable properties using quasi-realistic simulated environment of a possible future muon collider experiment

    The paper is structured as follows. Section~\ref{sec:ANewAlgorithm} introduces the new algorithm we propose.
    Section~\ref{sec:Simulation} details the detector used in our simulation and explains the simulation process.
    Section~\ref{sec:Reconstruction} outlines how the pattern recognition works to reconstruct tracks and reject background.   
    Section~\ref{sec:OptExecSpeed} and section~\ref{sec:HTAOptimization} discuss the optimization of the pattern recognition algorithm parameters to enhance execution speed.
    Section~\ref{sec:EventGeneration} describes the types of events simulated to test the pattern recognition algorithm. 
    Section~\ref{sec:Results} discusses the main results of the simulations.
    Section~\ref{sec:MassFits} demonstrates how the algorithm can be used to perform particle identification by adding the mass as an additional parameter of the fit.
    In section~\ref{sec:ExecutionTime}, we study the execution time of the algorithm as a function of the number of hits do be processed. Finally, we share our conclusions and outlook for the future in section~\ref{sec:Conclusions}.

  \section{A new algorithm}
  \label{sec:ANewAlgorithm}
    
    We propose to depart from the widely adopted class of Kalman filter based algorithms~\cite{Kalman} and explore the potential advantages of a new class of algorithms loosely inspired by the Hough transform concept~\cite{Hough}. One important goal is to try to reduce the time spent in making combinations of hits, which plague early stages of most currently adopted track reconstruction algorithms and cause the execution time to grow with some power of the hit density. This is problematic when the number of hits is very high.
    Ideally, we would like an algorithm whose execution time grows linearly with the number of hits. 
    We also would like the capability of easily adapting to different detector geometries. For example, we would like be able to provide a sample of simulated tracks  with the corresponding hits generated in a detector of arbitrary geometry and the code would {\em learn} how to perform pattern recognition and extract track parameters without the need to change the algorithm.
    We also would like to be able to include the time of arrival of the hit, if available, as an additional coordinate to be treated in the same way as a spatial coordinate, both in the pattern recognition stage (to find the track and discriminate it from background) and in track fitting stage (to extract track parameters, possibly including the mass of the particle).

    
    \begin{wrapfigure}{r}{0.4\textwidth}
        \centering
        \includegraphics[width=0.4\textwidth]{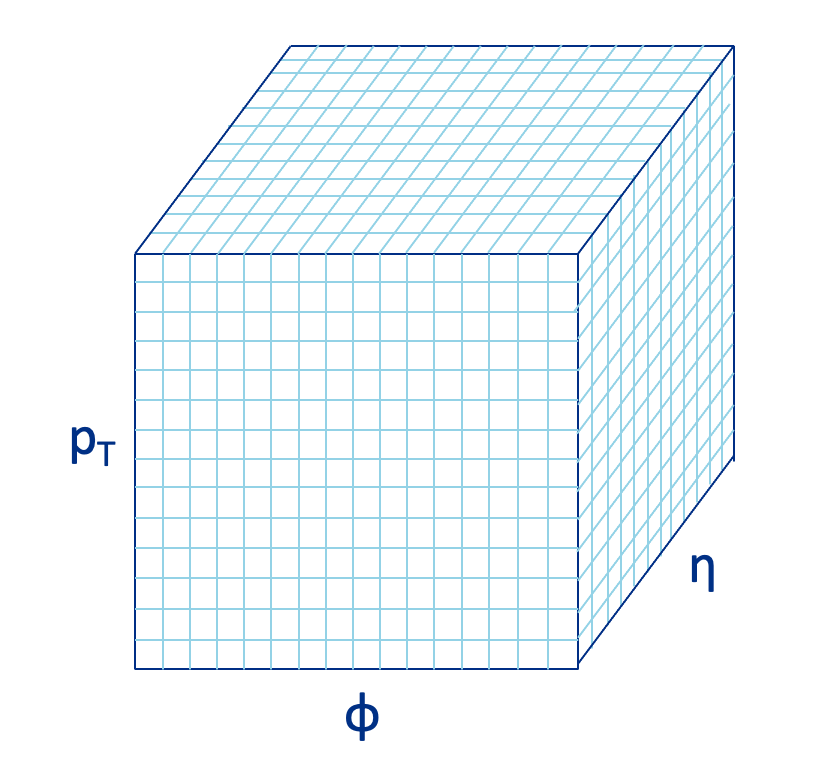}
        \caption{Conceptual illustration of the Hough Transform Array used in this study.
        \label{fig:Array}}
    \end{wrapfigure} 
    
    The pattern recognition algorithm we present here is based on ideas already discussed in reference~\cite{MDHT} and is based on a multidimensional extension of the Hough Transform.
    We refer to it as MDHT (Multi Dimensional Hough Transform).
    
    We begin by subdividing the parameter space of the tracks we want to reconstruct into a sufficient large number of small cells. In this way, we construct what we call the Hough Transform Array (HTA). Each dimension of the array corresponds to a different parameter of the tracks we want to reconstruct (e.g. $\phi, \eta, p_T$, as shown in figure~\ref{fig:Array}).
    The number of cells in the array needs to be optimized and this will be discussed in detail in section \ref{sec:HTAOptimization}.
    Then we simulate a sufficient number of tracks with parameters from each array cell, for each track we find the coordinates of the hits on all detector layers and record in a database the minimum and maximum value of each coordinate for each detector layer and each array cell.
    To perform pattern recognition, we will go through all the hits and assign each one of them to each cell in the array whose coordinate limits stored in the database for the correct layer are compatible with the actual coordinate values of the hit.
    Once all the hits have been processed this way, a good track candidate will show up as a cell with a sufficient number of hits assigned to it. 
  
    An important property of this algorithm is that the execution time is simply proportional to the total number of the hits to be processed, making it particularly attractive for high occupancy situations.
    The algorithm needs to be {\em trained} using a sufficiently large set of simulated tracks. The same track finding algorithm can be used for very different detector geometries and only the set of simulated tracks used for training needs to be changed. The particular structure of the algorithm also lends itself naturally to parallel hardware implementations, which, combined with its intrinsic flexibility, should provide a most powerful tool for triggering at future colliders.

\section{Simulation}
\label{sec:Simulation}

    To probe the effectiveness and performance of the algorithm, we decided to apply it to quasi-realistic simulated particle collision events.
    For this purpose, we chose to simulate the environment of a possible future muon collider~\cite{EPJC} because of its particularly challenging background environment for track reconstruction and because it is one of the most appealing developments for the future of experimental Elementary Particle Physics.
    
    The basic strategy is to choose a realistic, although approximate, geometry for the detector, simulate all the background hits that are expected to be produced in the detector, generate a single track within a certain parameter region of interest, simulate all the hits produced by that track in the detector and mix them with the background hits, submit the list of all the hits to the pattern recognition algorithm, and see how efficiently the hits coming from the track can be distinguished from the background hits. 
    The goal is to select enough {\em good} hits (produced by the track) in order to be able to perform a fit to the track trajectory and be able to extract the track parameters ($\phi, \eta, p_T$, etc) and reject {\em bad} hits (background hits) to avoid the fit to be distorted and the track parameters mis-measured.
    
    \subsection{The detector}
    
     \begin{figure}[t]
        \centering
        \includegraphics[width=1.0\textwidth]{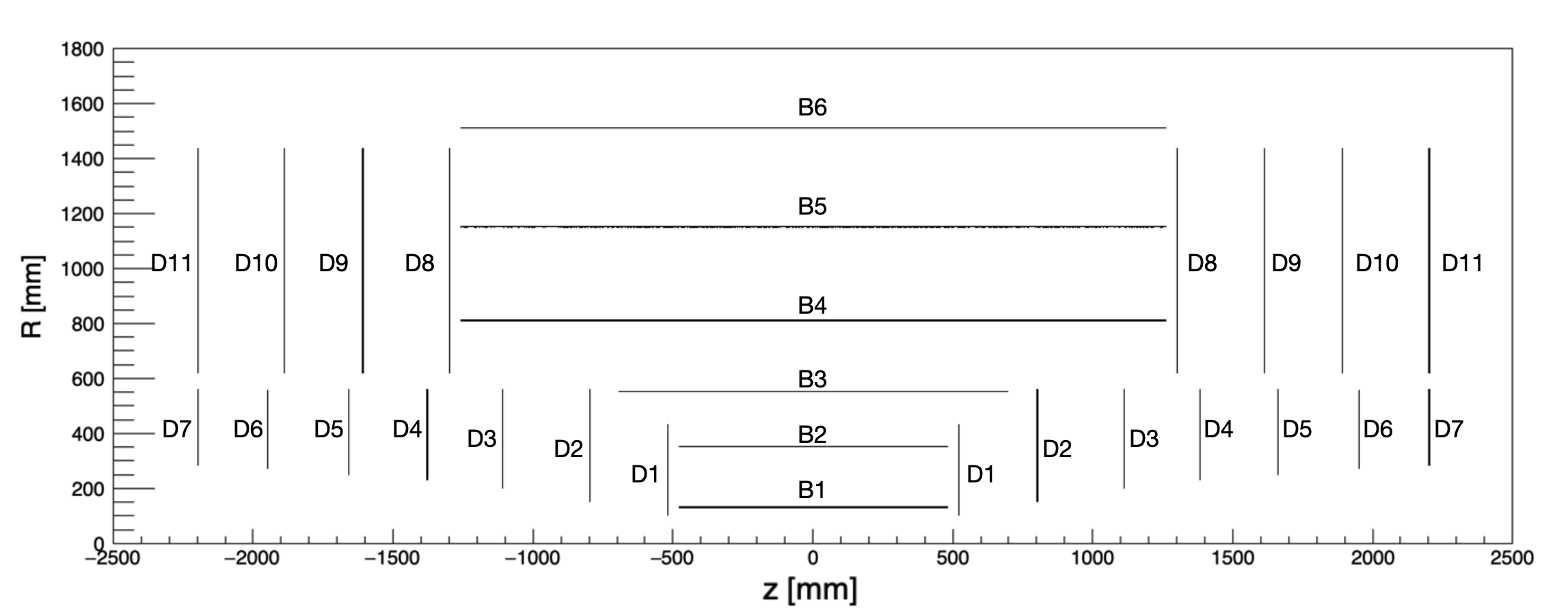}
        \caption{Layout of the simulated detector used in the study: the $R$-$z$ view is shown.}
        \label{fig:layout}
    \end{figure}
 
    The tracking detector simulated in these studies is based on a simplified version of the classical cylindrical layout typical of multipurpose collider experiments. We use a right-handed coordinate system with the origin located at the nominal interaction point. The $z$ axis is aligned with the direction of the clockwise-rotating positively charged beam. The $x$ axis lies on the collider plane, pointing away from the center of the collider ring, while the $y$-axis points upwards. We rely on a cylindrical coordinate system with $R$ being the radial distance from the $z$ axis, while $\theta$ and $\phi$ denote the polar and the azimuthal angle, respectively. Pseudo-rapidity is defined as $\eta=- \log\tan(\theta/2)$.
    
    The design of the detector stems from the muon collider detector, which is described in detail in section~4 of ref.~\cite{EPJC}. 
    The muon collider detector was in turn derived from the CLICdet concept developed by the CLIC collaboration~\cite{CLIC}. However, significant changes had to be introduced in order for the detector to be able to cope with the muon collider environment. Specifically, muon collider detectors are required to successfully disentangle the products of the $\mu^{+}\mu^{-}$ collisions from an intense beam-induced background (BIB) originating in interactions of the muon decay products with the machine components. For this reason, an essential part of the Machine Detector Interface (MDI) at a muon collider is a pair of cylindrical tungsten shields cladded with borated polyethylene (``nozzles'') placed along the beam pipe on both sides of the interaction point. These nozzles reduce the rate and energy of BIB particles that reach the detector by several orders of magnitude, however they also reduce the pseudo-rapidity coverage of the detector to about $|\eta|<2.4$.
        
    The tracking system geometry that we simulated for the scope of this work is shown schematically in figure~\ref{fig:layout}. The detector is an all-silicon tracker composed of barrels and disks, which are assumed to be infinitely thin and to have perfect cylindrical symmetry. 
    We believe that this gross simplification is still good enough to demonstrate, at least to first order, how this pattern recognition algorithm works and how it performs.
    
    Given the assumed perfect cylindrical symmetry around the $z$ axis, each horizontal line in figure~\ref{fig:layout} corresponds to a cylinder and each vertical line corresponds to a disk.
    The system is composed of two sub-detectors. The Inner Tracker consists of three barrel layers (B1-B3) and seven disks (D1-D7), covering a radial distance from approximately 150 mm to 550 mm. The Outer Tracker has three barrel layers (D4-D6) and four disks (D8-D11) and extends the radial coverage to 1500 mm. 
        
    The muon collider detector proposed in ref.~\cite{EPJC} also includes a Vertex Detector, positioned closest to the beamline and made of four layers of high spatial resolution silicon sensors. Due to its proximity to the interaction region, this sub-detector suffers from the highest hit density and the decision was made not to use it in the initial studies presented here. It is not shown in figure~\ref{fig:layout}. 
    Having said that, the tracking algorithm presented here can, at least conceptually, be easily extended to include hits from the Vertex Detector and we plan to incorporate them in the future.
        
    For barrels, we assume a position resolution of 10~$\upmu$m in the $\phi$ direction and 100$~\upmu$m in the $z$ direction. 
    For disks, we assume a position resolution of 10$~\upmu$m in the $\phi$ direction and 100~$\upmu$m in the radial direction.
    We also assume a 1\% hit inefficiency everywhere.
    The time resolution is assumed to be 60~ps everywhere and the whole tracker is immersed in a uniform 4.0~T magnetic field parallel to the $z$ axis.

    \subsection{Space-time tracking}

    We assume that, for every hit generated by a track passing through a detector layer, in addition to the measurement of two spatial coordinates, we also get a measurement of time.
    More precisely, we assume that we measure the time when the particle goes through the detector layer relative to the time of the beam crossing, as provided by the machine clock.
    Actually, since we are only interested in time differences, it is not so important what the reference of the time measurement is, as long as it is the same for all the hits.
    As stated above, time is assumed to be measured everywhere with a precision of 60~ps.

    One of the fundamental innovative design choices that characterize this algorithm is the fact that we want to treat the time coordinate of each hit in exactly the same way as the spatial coordinates. This is where the name {\em space-time tracking} comes from. We want to make no distinction between time and space both in the pattern recognition phase and in extraction of the parameters of the tracks (fitting phase). 
    As we will see, this allows us to use the information provided by the time measurement to the maximum extent possible and gives us the maximum discrimination power against the background.
    We will also see that fitting tracks in four dimensions allows us to perform some level of particle species identification by adding the mass of the particle as a parameter of the track as a simple extension of the fit.
    Since we want to treat space and time coordinates in the same way, we are also going to use millimeters to measure both lengths and times throughout the rest of this paper.
    So, for example, we will quote the time precision of our simulated detector as $\sigma_t = 18$~mm, which is equal to 60~ps times the speed of light.

    \subsection{Tracking simulation}

    We perform a simplified simulation of the path of charged particles through the detector.
    We assume a uniform magnetic field parallel to the $z$ axis of the detector.
    Therefore, we make the trajectory of each charged particle to be a perfect helix with its axis also parallel to the $z$ direction.
    
    Each hit is characterized by three coordinates $x_1$, $x_2$, $t$, where $x_1$ is the more precise spatial coordinate ($\phi$ direction for both barrels and disks), $x_2$ is the less precise spatial coordinate ($z$ for barrels and $R$ for disks), and $t$ is the time of arrival.

    The positions of all hits are determined first as pure geometrical intersections between the helix and the ideal shape of each detector (cylinder or disk) and then smeared with Gaussian errors.
    The time of arrival is calculated assuming constant speed along the helix starting from the time of origin and taking into account the momentum and the mass of the particle. The result so obtained is then also smeared with a Gaussian error.

    \subsection{Limitations of scope and disclaimer}
    
    For the purpose of this work, the generation of hits is the result of a pure geometrical intersection between an ideal track and an ideal detector shape.
    Effects of multiple scattering and other secondary interactions of particles with detector material are not accounted for.
    Also, effects due to the angle of impact of the track on the surface of the detector or due to charge sharing are ignored. The process of clustering is not simulated.
    Hits are not merged, even if they fall very close together.
    We believe that this crude level of approximation is good enough for the purpose of this work.
    
    Given all these limitations, the performance assumed for the detector in terms of position resolution are probably unrealistic, still we believe this is appropriate to demonstrate the capabilities of the algorithm.
    We do not want the limited performance of the simulated detector to obscure possible flaws of the reconstruction.
    For this reason, we need to keep in mind that we cannot attach any real significance to the results we obtain on the precision of the parameters we extract from the fits like, for example, momentum resolution.
    That would be outside the scope of this work and would require a much more realistic simulation of the detector, such as the one used in ref.~\cite{EPJC}.
    In this work, such results are presented with the sole purpose to prove to the reader that the algorithm is performing as expected.

    \subsection{Simulation of beam-induced background}  
    \label{sec:BIB}
        
    The beam-induced background at a muon collider is different from the machine background observed at current and past colliders. It is not produced by collisions in the interaction region, but is intrinsic to the muon beams.
    Muon decays occurring at distances of up to tens of meters from the interaction point can contribute to the background in the detector. Consequently, the simulation of the BIB is not trivial and requires a detailed and accurate modeling of the machine elements and the MDI. Since a fast parametric simulation of the BIB in the detector is not yet available, in this study we use a BIB sample that was generated by the U.S. Muon Accelerator Program~\cite{MAP} with the MARS15 code~\cite{MARS15} for a muon collider operating at a center-of-mass energy of 1.5 TeV. The sample contains approximately $10^{8}$ particles and corresponds to one crossing of the $\mu^+$ and $\mu^-$ beams, i.e. one event.
    The generated BIB sample was processed with a detailed detector simulation, based on \textsc{Geant4}~\cite{GEANT4}, and the reconstruction software~\cite{MuCollSoft1, MuCollSoft2} of the International Muon Collider Collaboration~\cite{IMCC} in order to produce a set of background tracker hits. 
    
    \begin{table}[t]
        \centering
        \caption{Position of the tracker layers with respect to the interaction point and average density of the beam-induced background hits. For the barrel layers ($BX$) and the endcap disks ($DX$), the radial distance $R$ and the the $|z|$ coordinate are provided, respectively.  
        \label{tab:hit_density}}        
        \medskip
        \begin{tabular}{|l c|c|c|}
        \hline
        Detector & Layer & $R/|z|$ [mm] & average hit density [mm$^{-2}$]   \\
        \hline
        Inner Tracker &  $B1$  &   129  &  0.738  \\     
        {}            &  $B2$  &   350  &  0.215  \\     
        {}            &  $B3$  &   550  &  0.095  \\     
        {}            &  $D1$  &   520  &  0.147  \\     
        {}            &  $D2$  &   800  &  0.081  \\     
        {}            &  $D3$  &  1110  &  0.058  \\     
        {}            &  $D4$  &  1380  &  0.043  \\     
        {}            &  $D5$  &  1660  &  0.031  \\     
        {}            &  $D6$  &  1950  &  0.025  \\     
        {}            &  $D7$  &  2200  &  0.020  \\
        \hline
        Outer Tracker &  $B4$  &   810  &  0.040  \\     
        {}            &  $B5$  &  1150  &  0.024  \\     
        {}            &  $B6$  &  1510  &  0.014  \\     
        {}            &  $D8$  &  1300  &  0.014  \\     
        {}            &  $D9$  &  1610  &  0.014  \\     
        {}            &  $D10$ &  1890  &  0.012  \\     
        {}            &  $D11$ &  2200  &  0.011  \\ 
        \hline
        \end{tabular}
    \end{table}

    The reconstructed hits of the BIB sample were produced according to a realistic tracker geometry, with flat rectangular silicon modules arranged to approximate cylinders and disks \cite{EPJC}. In order for the hits to be used in our ideal detector, it was necessary to adjust their positions to match our simplified geometry: the hits in the barrel layers were moved radially to the closest ideal cylinder surface, whereas the disk hits were shifted along the $z$ coordinate to the closest ideal disk. 
    Table~\ref{tab:hit_density} reports the average density of the beam-induced background hits in the barrel and disk layers of the Inner and Outer Trackers. 
    
    In the event generation, the BIB sample described above is employed as a pool of background hits, from which hits are drawn to populate the tracker layers according to the expected occupancy. 
    For every event, an appropriate number of hits is randomly extracted from the pool. The number of hits to be extracted is Poisson fluctuated around the average total number expected.
    Given that the spatial distribution of BIB hits in the tracker is approximately uniform in azimuthal angle $\phi$, and given the perfect $\phi$ symmetry of the simulated detector, BIB hits are additionally randomized by drawing a new random $\phi$ for each one of them. This {\em trick} yields a significant improvement in the statistical significance of the BIB sample at our disposal.

\section{The reconstruction algorithm}
\label{sec:Reconstruction}

    The MDHT algorithm consists of two phases. 
    The first phase is the {\em Training} phase and is described in section~\ref{subsec:training}. It is performed only once (in principle) to create the database that will be used during pattern recognition.
    The training phase will need to be repeated only if something is changed in the detector or in the parameter space of the tracks we want to reconstruct, or in the granularity of the HTA (see  section~\ref{sec:ANewAlgorithm}).
    The second phase is the {\em Pattern Recognition} phase and is described in section~\ref{subsec:recognition}.
    This is performed for every simulated event and its task is to use the database produced during the training phase to reject all the background hits, identify the good hits produced by the simulated track, and perform a fit to extract track parameters.

    A great deal of flexibility of this algorithm stems from the fact that, if the geometry of the detector changes, you only need to run again the training phase and the algorithm should be good to go again with no modification.
    We have not yet tested this feature extensively, but we plan to do so soon, as we believe it to be a very strong point, especially in the early stages of a new detector design.

    \subsection{Training phase}
    \label{subsec:training}
    
    \begin{figure}[t]
    \centering
        \includegraphics[width=0.95\textwidth]{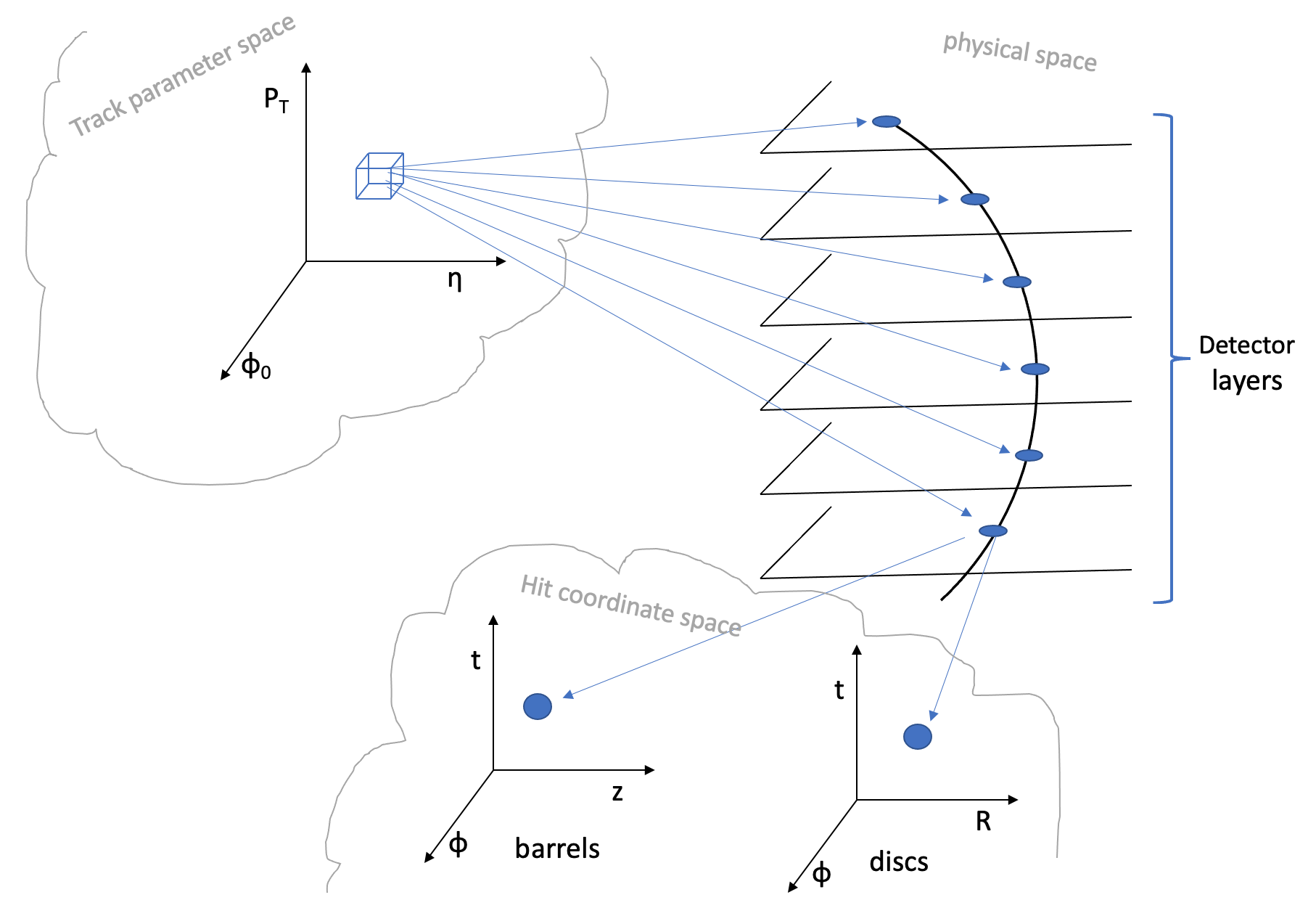}
        \caption{
            This figure illustrates the relationship between the track parameter space, the physical space, and the hit coordinate space.
            Given a small region in the track parameter space, we can simulate all the paths, in physical space, of all the tracks whose parameters are contained in that particular region.
            In each detector layer, barrel or disk, those tracks will generate a number of hits, which can be represented in a three-dimentional hit coordinate space (two space coortinates plus time).
            There will be a different hit coordinate space associated to each region in parameter space and each detector layer.          
        }
        \label{fig:ThreeSpaces}
    \end{figure}
    
    For each element of the Hough Transform Array, a fixed number of tracks is generated with parameters uniformly distributed inside the corresponding cell in parameter space. The number of tracks is the same for all the cells of the array and is currently equal to $500$.
    The path of each track through the detector is simulated and the coordinates of all the hits are obtained.
    Each hit is characterized by three coordinates $x_1$, $x_2$, $t$, as explained above, so that, for each HTA element and for each detector layer, we end up with a collection of three-dimensional points representing all the possible coordinates of the hits generated on that detector layer by the tracks coming from the parameter space cell corresponding to that HTA element.
    Obviously, only detector layers touched by that particular region of track parameter space will be present.
    For each one of these collections of hits, we need to find the minimum and maximum values of each coordinate and store them to be used  as boundaries during pattern recognition to decide if a particular hit may have been generated by a track coming from that region of parameter space.
    Finding the minimum and maximum values of the coordinates in each collection is a simple matter, but first we want to find the principal axes of the three-dimensional distribution and redefine new coordinates along those axes. This ensures that we take into account possible correlations between coordinates, minimizing the accepted volume in coordinate space and maximizing random noise rejection power.
    So we find the center of mass of each collection in three dimensions and define the three coordinates relative to the center of mass ($u_1$, $u_2$, $u_3$), along the principal components of the distribution.
    Given that $x_1$ runs along $\phi$ both in barrels and disks and given the perfect $\phi$ symmetry of our detector and track distribution, we only need to look for correlations between $x_2$ and $t$, so $u_1$ is the relative coordinate corresponding to $x_1$ and only the relative coordinates corresponding to ($x_2$, $t$) are rotated into ($u_2$, $u_3$) and only one rotation angle needs to be memorized.
    The relationship among the track parameter space, the physical space, and the detector coordinate space is schematically represented in figure~\ref{fig:ThreeSpaces}.

    \subsubsection{Structure of the HTA database}

    At the end of the training phase, a file is created with all the data that are needed by the pattern recognition algorithm.
    A list of detector layers is attached to each HTA element.
    This is the list of all the detector layers that are traversed by the tracks coming from the parameter region corresponding to that HTA element.
    For each detector layer, we need to store the center of gravity of the collection of hit coordinates (three parameters), to which the relative coordinates  $u_1$, $u_2$, $u_3$ will be calculated, the rotation angle (one parameter) of $u_2$, $u_3$ to the principal axes of the collection, and the minimum and maximum boundaries of $u_1$, $u_2$, $u_3$ after rotation (six parameters). 
    This file is the result of the training and is stored permanently. It is read out at the beginning of the pattern recognition phase and stored in memory for easy access by the pattern recognition algorithm.

    \subsection{Pattern recognition phase}   
    \label{subsec:recognition}

    In this section, we describe the actions  performed for every event in order to identify hits belonging to the same track and distinguish them from those belonging to the BIB. At this point, we assume that the data obtained during the training phase have been already loaded from the database into the Hough Transform Array.
    
        \subsubsection{Filling the Hough Transform Array}
        
        All hits are processed in sequence, one after the other. The processing order is irrelevant. In the simulation, all hits are randomly shuffled to eliminate any possible memory of the origin of each hit (track or BIB). 
        For each hit, the first task of the algorithm is to find all the cells in the HTA to which that hit can be {\em attached}. By definition, a hit must be {\em attached} to a particular cell if and only if that cell contains at least one combination of values of track parameters that describes a track compatible with having generated that particular hit. Given the work done during training, we assume this to be equivalent to saying that the coordinates of that hit are contained within the boundaries stored in the HTA database for that cell and for that detector layer.
        Actually, this is only an approximation, but it turns out to be more than adequate for our purposes.
        So, for each hit, we first find all the cells that have produced hits in the same layer as the layer this hit comes from, and then check if the coordinates of the hit are contained within the boundaries stored in the database. 
        To be more precise, we first translate the coordinates of the hit relative to the center of gravity and apply the rotation angle to the principal axes of the collection of the hit coordinates for that HTA element and that detector layer, then we check all six coordinate boundaries before we accept the hit.    
        We check all three coordinates $x_1$, $x_2$, $t$ and all three need to fall within the boundaries in order for the hit to be accepted and {\em attached} to that cell. In general, each hit will be attached to multiple cells, in fact there are infinite different values of the track parameters that are compatible with a single hit and those values typically span multiple cells.
        We call this process {\em Filling the HTA}.
        
        \subsubsection{Flagging the candidates}
        
        In order to have a chance to unambiguously determine the parameters of a track, we need to have hits on a number of different detector layers.
        So, for each HTA cell, we keep a count of the number of different detector layers populated by attached hits.
        When all the hits in one event have been processed, all the cells passing a given threshold in the number of different layers with hits are promoted to the status of {\em Candidate}. With this, we mean that there is a good chance that the hits attached to this cell include the hits of a good track whose parameters we are going to try to extract  through a minimum $\chi^2$ fit.
        At present, the minimum number of different detector layers populated by attached hits for a cell to be considered a {\em Candidate}, is five.
        
        \subsubsection{Fitting the candidates}
        
        All candidates in each event are considered one by one, and an effort is made to select the good hits, discard background hits, perform a minimum $\chi^2$ fit, and possibly extract track parameters from each one of them.
        We allow a maximum of one track to be extracted from each candidate.
        In case of ambiguity, we always choose the best solution on the basis of a {\em best fit} principle.
        The probability that two real tracks fall in the same HTA cell is very small and we neglect it. 
        In our simulation, that probability is actually zero since we generate only events with a single track.
        
                      
        The simplest possible situation is when the hits attached to a candidate are all on different detector layers.
        We will first describe what we do in this case.
        We construct a $\chi^2$ as a function of track parameters ($\phi$, $\eta$, $p_T$, $z_0$, $t_0$). The differences of the expected and the actual coordinate values ($x_1$, $x_2$, $t$) of each hit are divided by the measurement errors, squared and summed up for all the hits:
        \begin{displaymath}        
            \chi^2(\phi, \eta, p_T, z_0, t_0) = \sum_{\mathit{all\:hits}} \left[ \left(\frac{\Delta x_1}{\sigma_1}\right)^2 + \left(\frac{\Delta x_2}{\sigma_2}\right)^2 + \left(\frac{\Delta t}{\sigma_t}\right)^2\right] .
        \end{displaymath}
         $\Delta x_1$, $\Delta x_2$, $\Delta t$ are the differences of the coordinates of each hit with respect to those expected on the basis of the values of the track parameters in the arguments of the $\chi^2$ function.
         This function is minimized using the MINUIT algorithm~\cite{MINUIT} and parameters are extracted provided the value of the $\chi^2$ at the minimum is acceptable.
         This is a standard fitting procedure except that we are treating space and time coordinates exactly in the same way, hence our claim of ``{\em space-time tracking}''.
         
        \subsubsection{Solving ambiguities} 
        
        If multiple hits share the same layer, given that in the fit we use only one hit per layer, we are forced to examine different combinations of hits, fit them separately, and choose the one combination that we believe to be the most likely to include the good hits, that is those belonging to the real track and not coming from the background or extraneous tracks. 
        In choosing the best combination, we will be guided by a {\em goodness-of-fit} principle.
        
        At first, we combine all hits in all detector layers, so the number of combinations we obtain is equal to the product of the number of hits in each layer for this candidate. 
        If one or more of these combinations yields a fit with a good $\chi^2$, we take the parameters obtained from the fit with the best $\chi^2$ as the track resulting from this candidate.
        If none of these combinations yields a good $\chi^2$, then we try all combinations with one less hit, that is we eliminate one layer in turn from the combinations tried before. If we find some good fit, again we take the best $\chi^2$ fit as the track resulting from this candidate. If again we do not find any good fits, we eliminate yet another layer and we repeat this procedure until the number of layers left in the fit goes below the minimum allowed. At present, we allow a minimum of five layers to fit a track.
    
        \subsubsection{Dealing with duplicates} 

        Each hit is typically attached to multiple array cells. This happens because a single hit is, in general, not sufficient to pinpoint the
        parameters of the parent track to a single cell of the array.
        For this reason, it happens rather frequently that a single track creates multiple candidates, with exactly the same hits, in adjacent cells of the array. The fits of those candidates will yield exactly the same results leading to track duplicates, which must be explicitly removed.
        We do this by checking in which cell of the array the parameters of the track {\em resulting from the fit} fall, and allowing only one track per cell to be retained.
        All this will ensure that, in the end, we will have a maximum of one track inside each cell of the HTA and will reduce the probability of duplicates by orders of magnitude. 
        The potential loss of efficiency, due to multiple tracks being generated inside the same HTA cell is negligible due to the typically very large number of cells in the array.
        The computing time spent in fitting duplicates turns out to be negligible compared to the total execution times in the configurations that we have tested.

\section{Optimization of the execution speed} 
\label{sec:OptExecSpeed}

    \begin{figure}[t]
        \centering
        \includegraphics[width=0.70\textwidth]{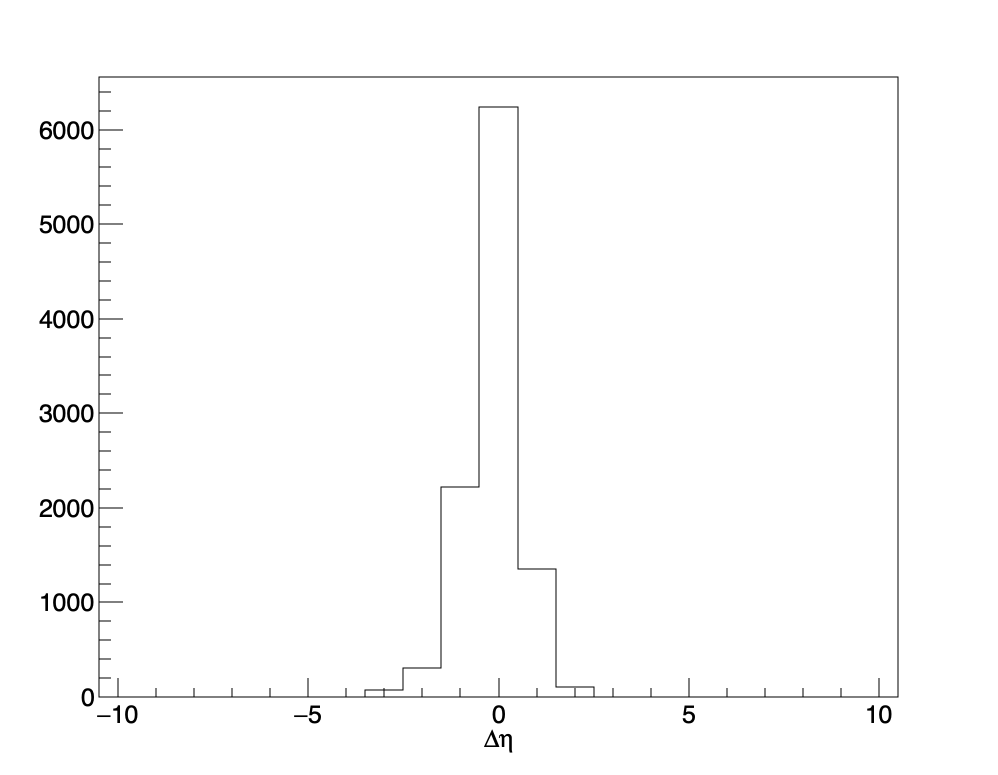}
        \caption{This plot shows the distribution, for all the hits in a test run, of the quantized distance in $\eta$ between all the cells a hit is attached to and the cell predicted only on the basis of the hit coordinates ($\Delta\eta$).
        The distance is expressed as the number of steps in the HTA.}
        \label{fig:delta_eta}
    \end{figure}

    \begin{figure}[p]
        \centering
        \includegraphics[width=0.70\textwidth]{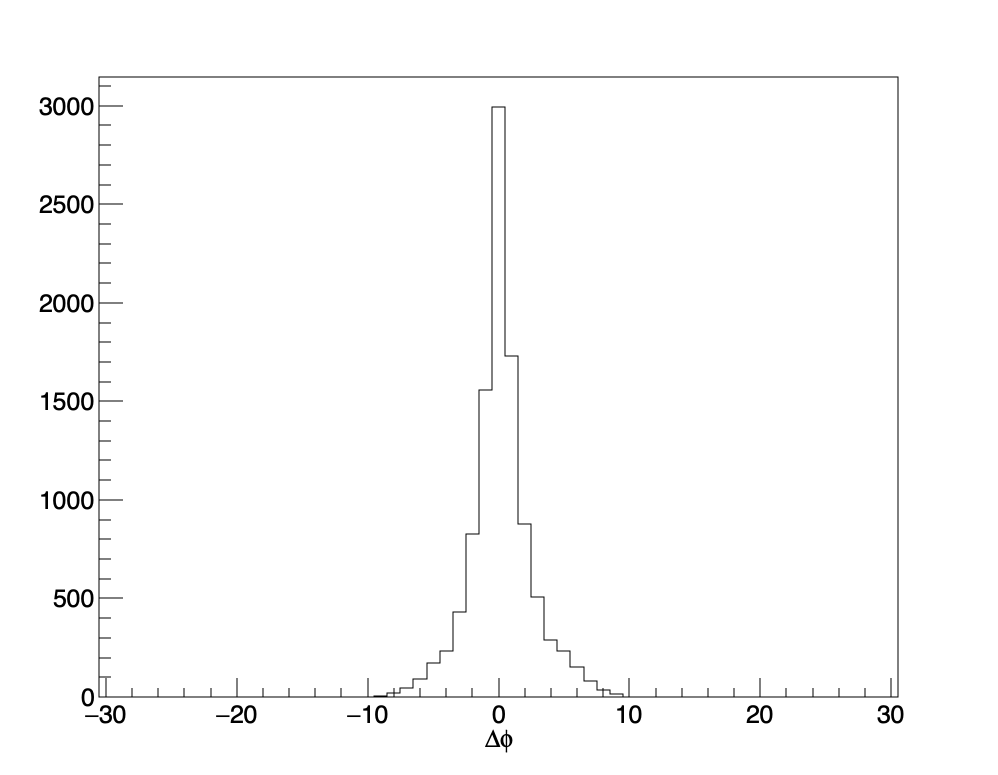} \\ \vspace{1cm}
        \includegraphics[width=0.70\textwidth]{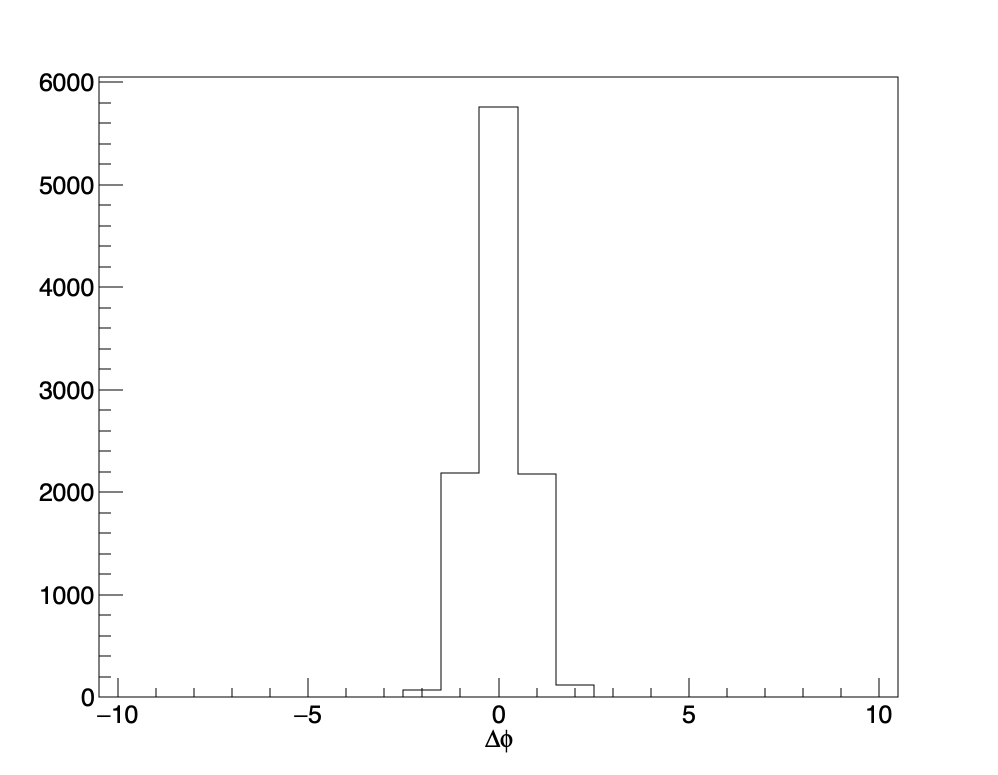}
        \caption{These plots show the distribution, for all the hits in a test run, of  the quantized distance in $\phi$ between all the cells a hit is attached to and the cell predicted only on the basis of the hit coordinates ($\Delta\phi$).
        The distance is expressed as the number of steps in the HTA. The plot on the top uses only the $\phi$ of the hit while the plot on the bottom includes a correction from the $p_T$ of the cell (note the different horizontal scales).}
        \label{fig:delta_phi}
    \end{figure}

    The execution time of the reconstruction algorithm is dominated by the time it takes to compare each hit to the cells of the HTA to find out which ones it needs to be attached to, and is therefore roughly proportional to the number of hits times the number of cells each hit is compared to.
    In principle, one should compare each hit with all the cells in the array, but that would be a huge waste of time because, just based on the values of the coordinates of the hit, one can exclude large regions of track parameter space, and the corresponding HTA cells.
    Having in mind the optimization of the pattern recognition algorithm execution time, we performed a study to find out 
    how much we can restrict the number of cells to be examined for each hit based only on its coordinates, without a significant loss of efficiency.
    For example, given the $z$ coordinate and the radius of the barrel, or the $R$ coordinate and the $z$ position of the disk, assuming the origin of the track to be at $z = 0$, we can estimate the $\eta$ of the track.
    In a calibration run, we plot the difference between the position in $\eta$ of each cell where a hit is attached by the algorithm, and the position predicted on the basis of the $\eta$ estimated from its coordinates only.
    An example of such a plot is shown in figure~\ref{fig:delta_eta}.
    The difference is an integer number because it is a difference between two indices of the HTA.
    From this plot, we can see with what accuracy we can predict which cells will be possible candidates for a given hit, based on its coordinates alone, and what margin of error $(\Delta\eta)$ we need to allow in order to maintain good efficiency.
    For each hit, we will compute the predicted $\eta_p$ and determine the range $[\eta_p-\Delta\eta, \:\eta_p+\Delta\eta]$ as the limited range in $\eta$ to be explored in the HTA for this hit.

    We can adopt a similar strategy for the parameter $\phi$.
    The additional challenge here is that also momentum comes into play in addition to hit coordinates. 
    Given the quantized momentum parameter and  the hit coordinates, we can estimate the quantized parameter $\phi_p$.
    Figure~\ref{fig:delta_phi} shows plots of $\Delta\phi$ defined in a similar way as $\Delta\eta$. The left plot is without momentum correction while the right plot is with momentum correction.
    From this plot we obtain $\Delta\phi$  and we restrict the $\phi$ range explored in the HTA to $[\phi_p-\Delta\phi, \:\phi_p+\Delta\phi]$.

    In our case, the application of this optimization to the HTA scan reduces the number of cells examined by the algorithm per hit and the total execution time by a factor of about 600.

\section{Optimization of the HTA dimensions}
\label{sec:HTAOptimization}

    \begin{table}[t]
        \centering
        \caption{For each HTA configuration, the table presents the number of $\phi$, $\eta$, and $p_T$ bins, $\Delta\eta$ and $\Delta\phi$ as defined in section~\ref{sec:OptExecSpeed}, the track-finding efficiency for single-track events, the number of candidates per event and the average processing times for the HTA pattern recognition without and with candidate fitting in the case of background-only events. The HTA spans the track parameter space defined by $0^\circ < \phi < 30^\circ$, $|\eta| < 2.5$, and $p_T > 3$ GeV/c.
        \label{tab:HTA_opt}}
        \medskip
        \begin{tabular}{|ccc|cc|c|c|c|c|}
        \hline
        \multicolumn{5}{|c|}{\textsc{HTA configuration}} & \multicolumn{4}{|c|}{\textsc{HTA performance}} \\
        $N_\phi$  & $N_\eta$  &  $N_{p_{T}}$ & $\Delta\eta$ & $\Delta\phi$ & $\epsilon_{\mathrm{trk}}$  &  $N_{\mathrm{\ cand}}^{\mathrm{}}$ & $T^{\mathrm{}}_{\mathrm{HTA}}$ [s] & $T^{\mathrm{}}_{\mathrm{HTA\, +\, fit}}$ [s] \\ 
        \hline
        10  &  540  &   6  & 2 & 1 & 0.961  &    38  &  16.3   &  16.9  \\
        12  &  450  &   6  & 1 & 1 & 0.963  &    34  &  10.2   &  10.4  \\
        15  &  360  &   6  & 1 & 1 & 0.963  &    36  &  10.1   &  10.2  \\
        15  &  180  &  12  & 1 & 1 & 0.963  &    65  &  19.2   &  19.6  \\
        15  &   90  &  24  & 1 & 1 & 0.963  &   248  &  36.2   &  37.9  \\
        30  &  180  &   6  & 1 & 2 & 0.963  &    80  &  15.0   &  14.9  \\
        30  &   90  &  12  & 1 & 1 & 0.963  &   149  &  18.5   &  19.0  \\
        30  &   45  &  24  & 1 & 1 & 0.963  &   737  &  36.1   &  41.2  \\
        60  &   90  &   6  & 1 & 3 & 0.963  &   282  &  19.6   &  20.4  \\
        60  &   45  &  12  & 1 & 2 & 0.963  &   792  &  28.1   &  32.3  \\
        120  &  45  &   6  & 1 & 7 & 0.962  &  2330  &  39.0   &  61.6  \\
        \hline
        \end{tabular}
    \end{table}
    The definition of the HTA dimensions depends on many factors and is not easy to determine {\it a priori}, since there are contrasting needs. 
    For example, it is rather intuitive that a finer granularity will provide higher resolution and better rejection against background hits, but will occupy more memory and, more importantly, will require more time for the hits to be compared to all the cells. 
    It is also difficult to predict how different {\em aspect ratios} of the HTA, i.e. different segmentations along different axes for a constant total number of cells, will affect performance. 
    Currently, in the absence of a guiding principle, a workable solution must be found through trial and error.
    
    Our studies focused on a track parameter space defined by $0^\circ < \phi < 30^\circ$, $|\eta| < 2.5$, and $p_T > 3$ GeV/c, and on the corresponding $\phi$ sector of the tracker spanned by all the hits produced by those tracks.
    In order to identify an appropriate HTA configuration, we started with an arbitrary array of 30 bins in $\phi$, 90 bins in $\eta$, and 12 bins in $p_T$, resulting in a total of 32,400 cells. We then varied the segmentation along $\phi$, $\eta$, and $p_T$, keeping the total number of cells constant, and for each HTA configuration we compared the total track-finding efficiency, the average number of candidates we need to fit per event, and the average time required to process an event with only background hits. 
    As a figure of merit, the objective was to minimize the processing time while maintaining high tracking efficiency and a reasonably small number of candidates to process.
    
    The results for the different HTA configurations tested are summarized in table~\ref{tab:HTA_opt}.
    The track finding efficiency was evaluated in a sample with a single track per event. The efficiency is less than 100\% because the geometrical acceptance of the detector ($|\eta_{\mathrm{det}}|<2.436$) is smaller than the $\eta$ range used in the track generation ($|\eta|<2.5$). Additional inefficiencies of the order of a few per mil are due to the $\Delta\eta$ and $\Delta\phi$ cuts.
    The number of candidates per event and the average processing times refer to background-only events.
    The processing times were estimated running a single process on an Intel\textsuperscript{\textregistered} Xeon\textsuperscript{\textregistered} W-2295 CPU @ 3.00GHz with 128GB RAM.
    The HTA performance is mainly driven by the array granularity in $\eta$.
    The execution time increases approximately linearly with $N_{p_{T}}$ because no HTA scan optimization was applied for $p_T$.

    We finally chose the configuration with $N_\phi = 15$,  $N_\eta = 360$, and $N_{p_{T}} = 6$, which will be fully characterized in the following sections.

\section{Event generation} 
\label{sec:EventGeneration}
    
    For each simulated event, we generate a single charged track, calculate its trajectory through the detector and obtain the coordinates of all the hits it produces.
    To this we add an appropriate number of random background hits in order to approximate as closely as possible the hit density and the distribution that are expected at the muon collider from the beam-induced background. 
    Section~\ref{sec:BIB} describes the procedure we use to simulate this background.
    
    Tracks are generated following a uniform distribution of the azimuthal angle at the origin within the interval $\phi_0 = [0^\circ, 30^\circ]$,
    a uniform distribution in pseudo-rapidity within the interval $\eta = [-2.5, +2.5]$, and a uniform distribution in inverse transverse momentum in the interval $1/p_T =[-1/3, +1/3]$ GeV/$c^{-1}$.
    We use the convention of assigning negative $p_T$ to negative charged particles.
    The position of the origin of the track is assumed to be exactly on the $z$ axis ($x=y=0$).
    The position along the $z$ axis is assumed to be distributed as a Gaussian centered at $0$ and with $\sigma_z = 1.5$~mm.
    The time of the origin of the track ($t_0$) is also assumed to be distributed as a Gaussian centered at $0$ and with $\sigma_t = 1.5$~mm/$c$ (5~ps).

\section{Results}
\label{sec:Results}

    \subsection[$\chi^2$ distribution]{$\bm{\chi^2}$ distribution}
    \label{sec:chi2}

    \begin{figure}[t]
     \centering
     \includegraphics[width=0.70\textwidth]{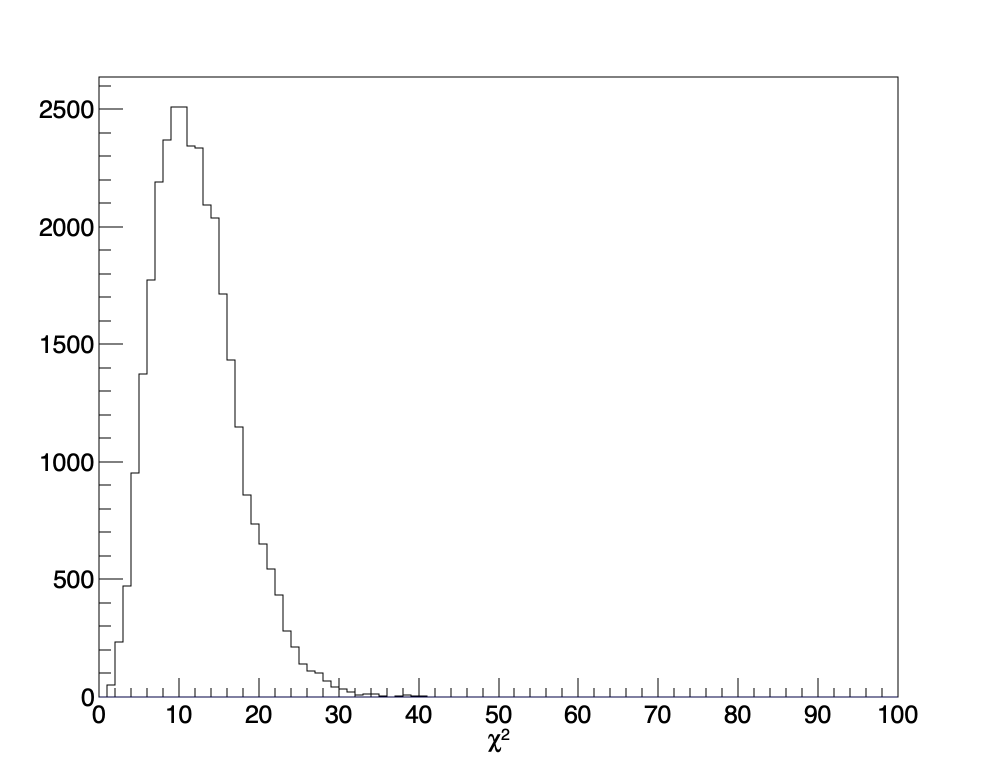}
     \caption{Distribution of $\chi^2$ resulting from all track fits.}
     \label{fig:Chi2}
    \end{figure}
    
    The distribution of the $\chi^2$ resulting from all track fits is shown in figure~\ref{fig:Chi2}.
    The distribution is very clean and does not show any visible feed-down from the high-$\chi^2$ fits
    produced by tracks contaminated by noise hits.
    The mean value of the distribution is what is expected from the mean number of degrees of freedom of the fits,
    typically 5-7 hits, three coordinates per hit for a five parameter fit ($\phi$, $\eta$, $p_T$, $z_0$, $t_0$).

    \subsection{Efficiency}

    \begin{figure}[p]
        \centering
        \includegraphics[width=0.70\textwidth]{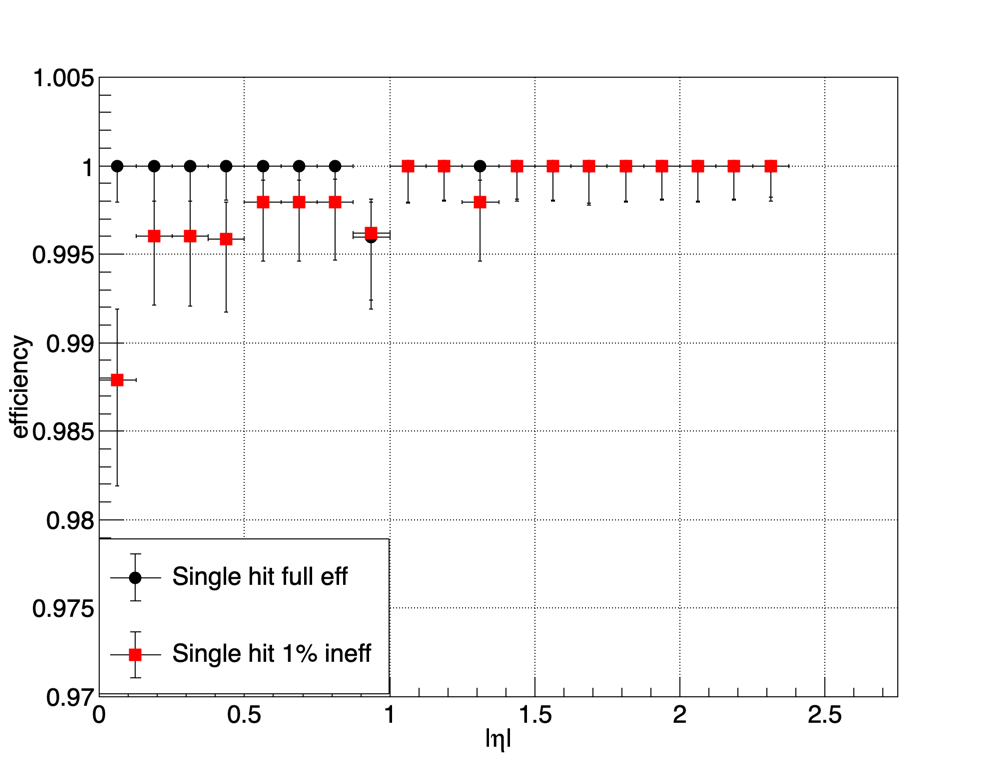} \\ \vspace{1cm}
        \includegraphics[width=0.70\textwidth]{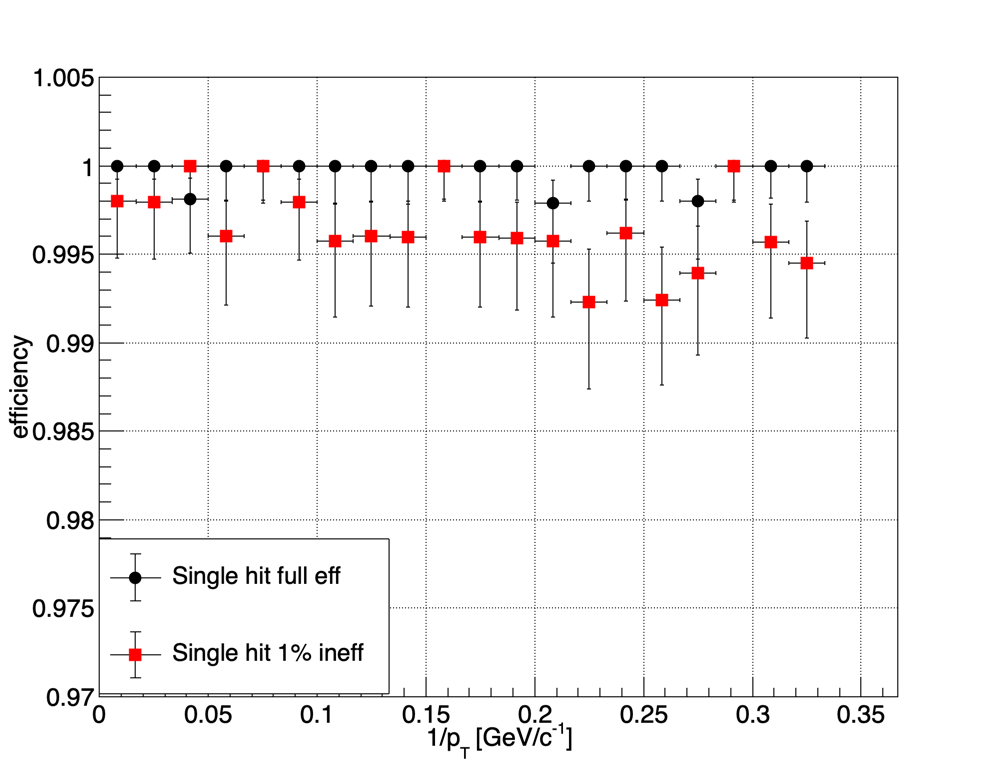}
        \caption{Tracking efficiency vs $|\eta|$ (top) and $p_T^{-1}$ (bottom).
        Black round markers show results obtained assuming 100\% single hit efficiency,
        while red square markers assume 1\% single hit inefficiency everywhere in the detector.}
        \label{fig:trk_eff_vs_eta_and_pt}
    \end{figure}

\begin{figure}[t]
        \centering
        \includegraphics[width=0.70\textwidth]{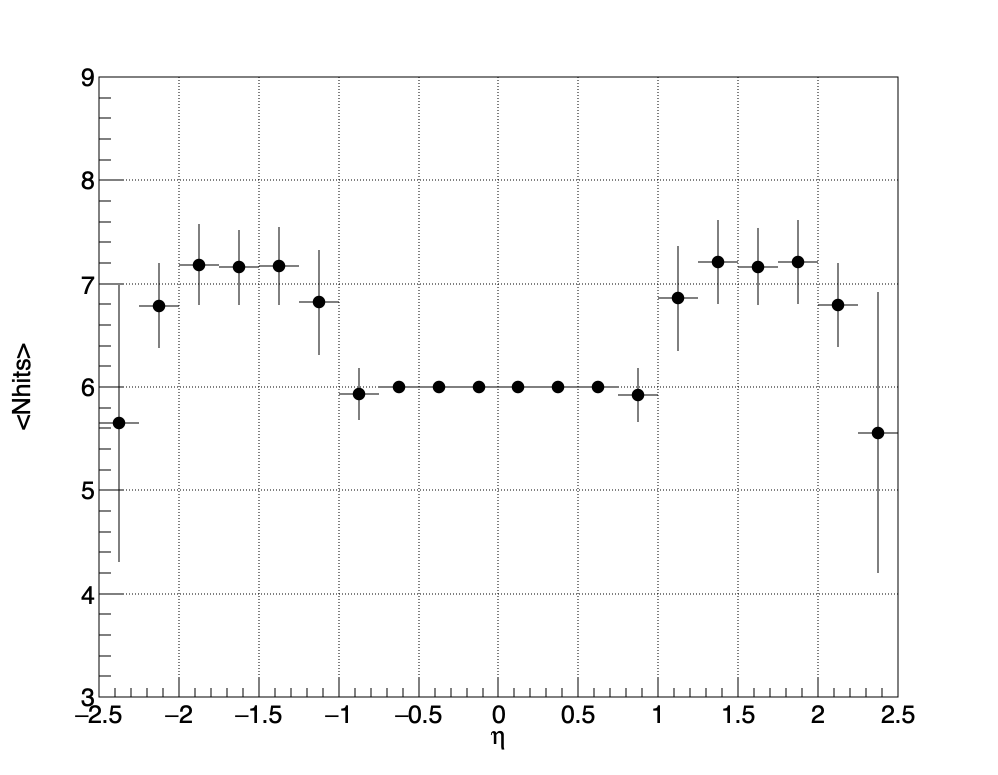}
        \caption{Average number of hits versus $\eta$ of the track.}
        \label{fig:Nhits_vs_eta}
    \end{figure}

    In our simulation, we generate one single track per event superimposed to an appropriate amount of hits from the BIB.
    To measure the efficiency of our tracking algorithm, we count the number of events where a track is found and the fit yields a good $\chi^2$.  We define the {\em efficiency} as the ratio of this number over the total number of events. 
    Moreover, we compute this efficiency separately for different regions of the parameter space where the original track is generated.
    For example, the tracking efficiency that we obtain as a function of pseudo-rapidity ($\eta$) and transverse momentum ($1/p_T$) is shown in figure~\ref{fig:trk_eff_vs_eta_and_pt}.
    The plots show the efficiency for two different assumptions for the detector single-hit efficiency. Black round markers correspond to the assumption of full (100\%) singe-hit efficiency while red square markers correspond to the assumption of 1\% single-hit inefficiency everywhere in the detector.
    The efficiency is extremely good everywhere and shows the capability of the algorithm to find the good hits even in the extremely demanding background environment expected at the muon collider.
    We can observe a minor loss of efficiency in the central region, near $\eta = 0$, and for lower transverse momenta. In fact, that corresponds to the region in parameter space where the track, given the geometry of the detector, gets the minimum average number of hits and is therefore most vulnerable to possible losses of the pattern recognition algorithm.
    This is shown, for example, in figure~\ref{fig:Nhits_vs_eta}, where the average number of hits produced is plotted as a function of $\eta$ of the track.
    We can see that the average number of hits goes from about six to about seven when moving out from the central region 
    $\eta = [-1, +1]$, to fall down again only for $|\eta| > 2$.
    This also suggests that the geometry of this detector might be optimized further by equalizing the coverage as a function of $\eta$.

    \begin{figure}[p!]
        \centering
        \includegraphics[width=0.70\textwidth]{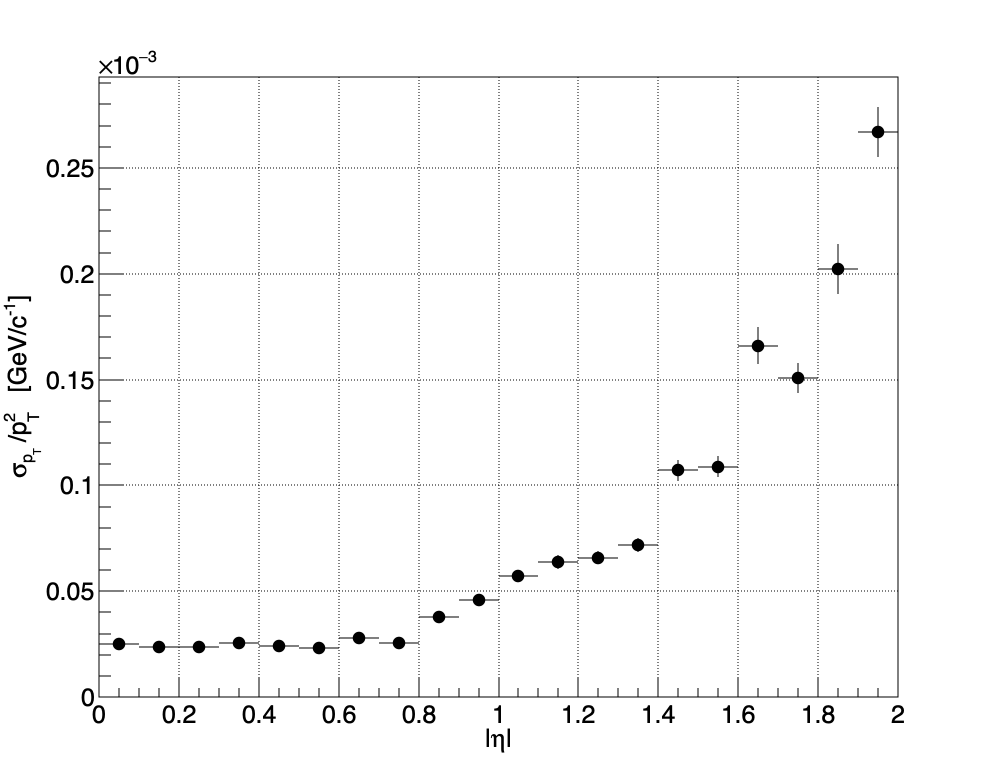} \\ \vspace{1cm}
        \includegraphics[width=0.70\textwidth]{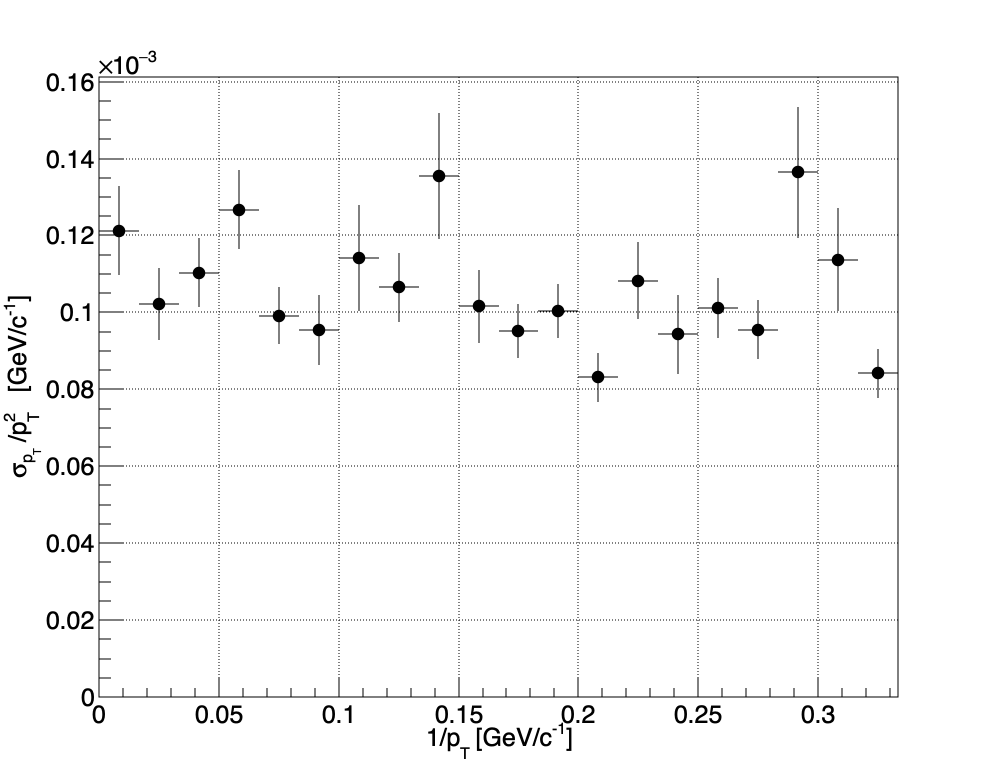}
        \caption{Momentum resolution $\sigma_{p_T}/{p_T}^2$ versus $|\eta|$ (top) and $p_T^{-1}$ (bottom).}
        \label{fig:mom_res_vs_eta_and_pt}
    \end{figure}    

      \subsection{Resolutions}

      For those events where a track is found with a good $\chi^2$, we can examine the parameters returned by the fit, compare them with the {\em true} parameters of the original generated track, and evaluate the standard deviation of the difference.
      For example, the momentum resolution that we obtain as a function of pseudo-rapidity ($\eta$) and transverse momentum ($1/p_T$) is shown in figure~\ref{fig:mom_res_vs_eta_and_pt}.
      Resolution is expressed as $\sigma_{p_T}/p_T^2$, as usual.
      This quantity is expected to be approximately independent of momentum if the effect of multiple scattering is negligible.
      Momentum resolution is in the vicinity of $2\times 10^{-5}$ GeV/$c^{-1}$ in the central region ($\eta=[-0.8,+0.8]$) and takes off rapidly going forward.
      Instead, it shows no apparent variation as a function of $p_T$ within statistical fluctuations (as expected in the absence of multiple scattering).

    \subsection{Effect of noise}

    To quantify the effect of the background on the results shown above, we ran the same simulated events with tracks only and no background hits. We produced the same plots but we could not see any difference whatsoever in either efficiency or resolution.
    We then inspected all the tracks that passed the $\chi^2$ cut and found that in 10,000 simulated events, out of a total of 56,737 hits, only 12 hits were coming from BIB, which is a fraction of about $2.1\times 10^{-4}$, so it is not surprising that they had no visible effect on the results.
    

    Such a good performance shows that, with the extremely good spatial resolution that we are assuming for the detector, even the high-density background expected at the muon collider can be easily discriminated, and the MDHT algorithm is performing well from that point of view.
    But assuming 10~$\upmu$m resolution is probably unrealistic, especially at low momentum, since we are not simulating multiple scattering.
    So we wanted to estimate what the effect of multiple scattering might be for a 3 GeV/$c$ particle.
    We assumed 0.1 radiation lengths for the total thickness of the detector, according to what was estimated in figure~36 of ref.~\cite{EPJC}, calculated the effect of multiple scattering on each detector layer for each simulated track using the usual small angle approximation (see eq.~34.16 in section~34.3 of ref.~\cite{PDG}), and took the average over all the tracks for each individual detector layer.
    The result was added in quadrature to the intrinsic resolution assumed for the detector and used as a new updated resolution independent of momentum and angle of incidence.
    This is a very crude approximation, but gives us an idea of the order of magnitude of the effect of multiple scattering and how the algorithm might be affected.
    The new updated resolutions typically ranged from 100 $\upmu$m to a few mm going from the innermost to the outermost layers.
    The fraction of background hits found in accepted tracks went up from $2.1\times 10^{-4}$ to $6.0\times 10^{-2}$, a very dramatic effect.
    The momentum resolution was degraded by about a factor 100. 
    So, even with these very large, pessimistic resolutions, the MDHT algorithm still performs  well.
    For example, the momentum resolution is what is expected given the detector spatial resolution, the lever arm, and the intensity of the magnetic field, and is compatible with the results shown in figure~39 of ref.~\cite{EPJC} for the momentum range we are considering here.

    
    \subsection{Fake track rate}
    \label{sec:fakes}

    To see what is the probability of the MDHT algorithm reconstructing a fake track from background only, we ran ten thousand simulated events containing BIB hits only.
    In the version where we assume only intrinsic detector resolutions (10 and 100 $\upmu$m in orthogonal directions) and ignore multiple scattering, we observe no track passing the $\chi^2$ cut.
    In the version where we use expanded resolutions (100 $\upmu$m to a few millimeters depending on which layer), we observe 576 tracks passing the $\chi^2$ cut in ten thousand events. 
    If we assume that the probability of mistakenly accepting one random BIB hit as a good hit is directly proportional to coordinate measurement errors, then we expect the probability of making a seemingly good track out of, say, five hits on five different layers to be proportional to the product of the fifteen resolutions of the fifteen coordinates involved (three per layer, including time).
    If we take the geometrical mean of all degradation factors (expanded resolution divided by intrinsic resolution) of all coordinates of all detector layers in the central region, where almost all the fake tracks are found, we obtain an average degradation factor of 7.1. 
    For a fit with five hits and ten degrees of freedom, we expect an increase in fake rate of about $7.1^{10} = 3.3\times10^{8}$.
    So it is not surprising that we observe no fake tracks when we assume nominal resolutions; in fact, if we scale down 576 by $3.3\times10^{8}$, we obtain $1.7\times10^{-6}$ expected tracks per ten thousand events.

    \begin{figure}[p]
        \centering
        \includegraphics[width= 0.9\textwidth]{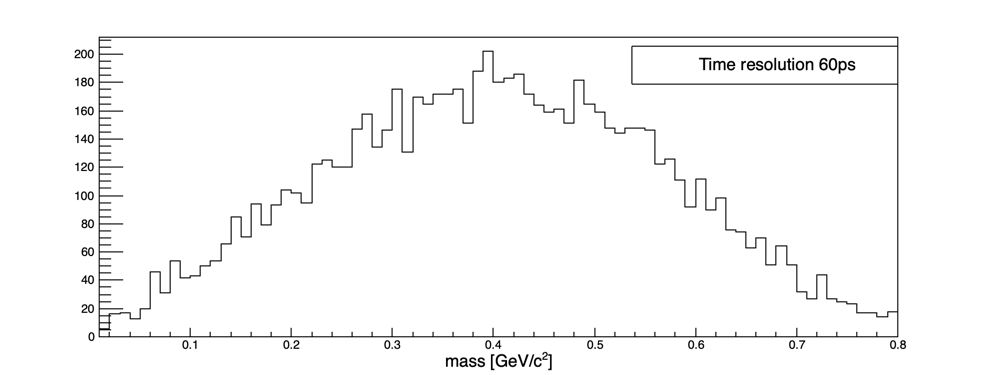} \\ \vspace{1cm}
        \includegraphics[width= 0.9\textwidth]{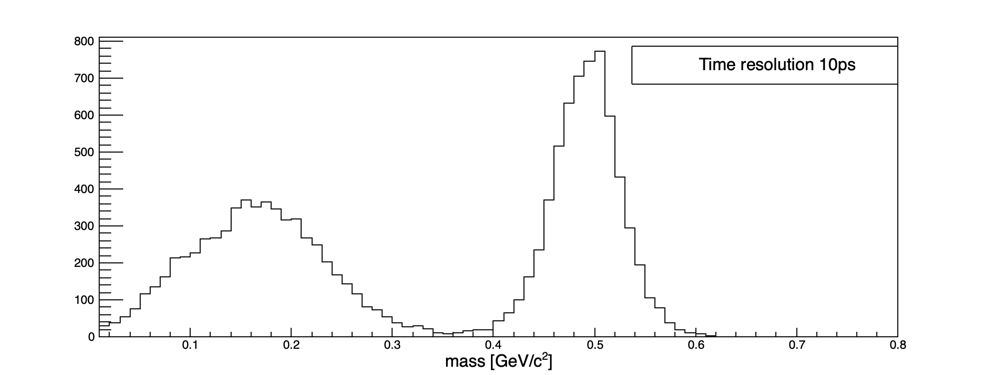} \\ \vspace{1cm}
        \includegraphics[width= 0.9\textwidth]{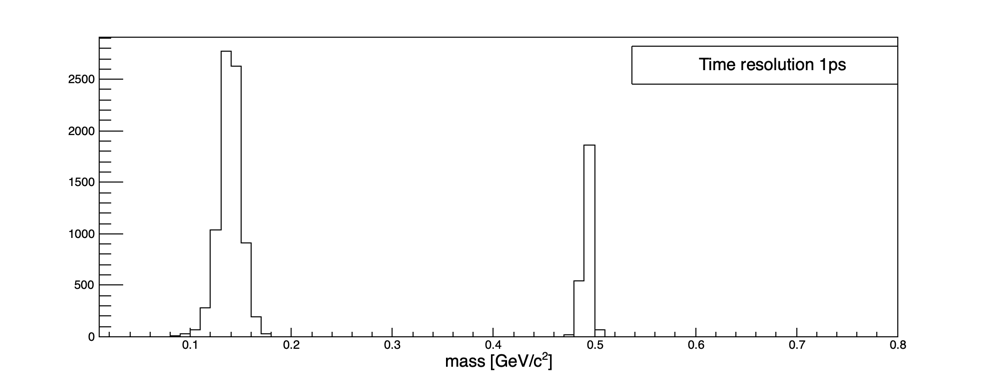} \\
        \caption{Distributions of the mass of the track as obtained when its value is treated as a free parameter in the fit. The simulated sample is a mixture of charged pions and kaons. Only tracks with momentum below 3 GeV/$c$ have been selected.  The time resolution of the detector has been assumed to be 60 ps (top), 10 ps (middle), 1 ps (bottom).}
        \label{fig:mass_plots}
    \end{figure}

\section{Fitting for particle mass}
\label{sec:MassFits}

    We assume our detector to be able to measure, in addition to the spatial coordinates, also the time of arrival of each hit. This means that we can measure the time of flight of each particle and, in principle, we can identify different particle species, provided we have sufficient time resolution and provided the momentum of the particles is sufficiently low.
    In practice, we can do this by simply adding the mass as an additional track parameter to the fit of each track candidate.
    The $\chi^2$ function provided to MINUIT will then be a six-parameter function:    
     \begin{displaymath}        
            \chi^2(\phi, \eta, p_T, z_0, t_0) \longrightarrow  \chi^2(\phi, \eta, p_T, z_0, t_0, m)\ . 
        \end{displaymath}
    The fit will return the best guess for the particle mass in the same way as all the other parameters.
    To test this idea, we generated events with a mixture of pions and kaons with transverse momentum between 1.5 GeV/$c$ and 3.0 GeV/$c$ and performed a mass fit to all tracks with absolute value of momentum below 3.0 GeV/$c$. Mass plots are shown in figure~\ref{fig:mass_plots} assuming different time resolutions.

\section{Execution time}
\label{sec:ExecutionTime}

    \begin{table}[t]
        \centering
        \caption{Number of candidates and average execution times per event for pattern recognition alone and pattern recognition plus the track fitting for different hit densities in the detector. The factor in the first column multiplies the BIB hit densities reported in table~\ref{tab:hit_density} per tracker layer.
        The results refer to the HTA configuration with $N_\phi=15$, $N_\eta = 360$, and  $N_{p_{T}}=6$.
        \label{tab:BIB_frac}}
        \medskip
        \begin{tabular}{|c|c|c|c|}
        \hline
        BIB multiplier & $N_{\mathrm{cand}}^{\mathrm{}}$ & $T^{\mathrm{}}_{\mathrm{HTA}}$ [s] & $T^{\mathrm{}}_{\mathrm{HTA\, +\, fit}}$ [s] \\
        \hline
        0.5  &     2  &   5.1  &    5.1  \\
        1.0  &    36  &  10.1  &   10.2  \\
        1.5  &   123  &  15.4  &   16.0  \\
        2.0  &   283  &  20.5  &   23.2  \\
        2.5  &   499  &  25.9  &   33.8  \\
        3.0  &   785  &  31.2  &   51.5  \\
        3.5  &  1104  &  36.3  &   80.9  \\
        4.0  &  1457  &  41.8  &  127.3  \\
        4.5  &  1854  &  47.5  &  205.1  \\
        5.0  &  2273  &  52.3  &  325.2  \\
        \hline 
        \end{tabular}
    \end{table}

    The execution time of the MDHT algorithm is expected to be proportional to the number of hits to be processed. This is a particularly desirable and advantageous feature in cases where hit densities are high and may have large fluctuations. 
    
\begin{figure}[t]
        \centering
        \includegraphics[width=0.7\textwidth]{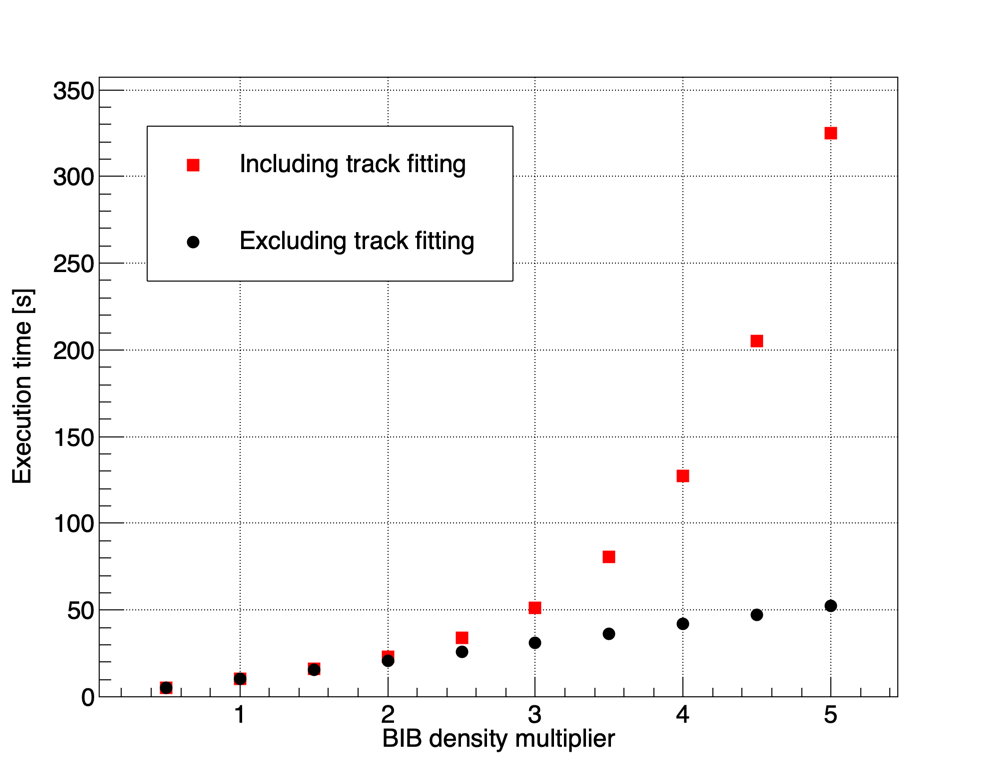}
        \caption{MDHT execution time for the pattern-recognition stage and the pattern recognition plus fitting stage as a function of the hit density in the tracker.
        \label{fig:BIB_frac}}
    \end{figure}

    The timing performance of the algorithm has been studied using the background sample.
     To assess the impact of different hit densities on the MDHT algorithm timing, we evaluated the execution time per event while gradually increasing the BIB density by factors of up to 5 in the case of our reference HTA configuration with $N_\phi=15$, $N_\eta = 360$, and  $N_{p_{T}}=6$. For each hit density scenario, we estimated the execution time required for both pattern recognition and pattern recognition followed by track fitting by running a single process on an Intel\textsuperscript{\textregistered} Xeon\textsuperscript{\textregistered} W-2295 CPU @ 3.00GHz with 128GB RAM.
    The expected number of candidates per event and the pattern-recognition and total execution times per event are presented in table~\ref{tab:BIB_frac}.
    The execution times are also shown in figure~\ref{fig:BIB_frac} as a function of the BIB multiplier.

    We observe a very good linearity of the total processing time up to a factor of about 2.5, after which a non-linear take-off occurs. If we remove the fitting stage and leave only the pattern recognition phase, the linearity is restored.    
    Non-linearity is due to the large number of hit combinations to be fitted and could be at least partially corrected by increasing the HTA granularity. For example, in the case of a BIB multiplier of five, with doubled granularity of $N_\phi=30$, $N_\eta = 360$,  $N_{p_{T}}=12$, the number of candidates is reduced from 2273 to 1935, the time for performing pattern recognition grows from 52.3 s to 120.1 s due mostly to the doubling of the number of cells in the HTA, but the total processing time, including the time needed to fit all the candidates, goes down from 325.2 s to 150.8~s. 
    This means that the HTA configuration we are using is not optimal, in terms of execution time, for BIB densities that high.
    It is intuitive that higher BIB densities will demand higher HTA granularity. This means that, if BIB density is five times what we have assumed for the optimization of the HTA configuration, then we should perform a new optimization assuming the correct BIB density and we will end up with a higher granularity configuration and a restored linearity in the vicinity of the correct BIB density.

\section{Conclusions}
\label{sec:Conclusions}
    
    We believe we have shown that the Multi Dimensional Hough Transform (MDHT) tracking algorithm described in this paper
    is expected to perform  well in terms of rejecting background even in the very challenging environment foreseen at future
    colliders and, in particular, at a muon collider.
    The execution time being linearly proportional to the number of hits to be processed is a very desirable feature in the quest for the best possible time performance and stability versus possibly unexpected intensity fluctuations.
    The inclusion of the time of arrival as an additional coordinate to be treated in the same way as the spatial coordinates (whence the {\em space-time tracking} claim) leads to the full exploitation of the information carried by the time measurement both for the discrimination from background and in the determination of track parameters. This will be particularly important if the accuracy of time measurement will improve in the years to come due to multiple ongoing R\&D efforts.
    This also offers the possibility of a new approach to particle identification by opening the possibility of obtaining the mass of a particle as an additional parameter returned by the fit of a track.
    
    The particular architecture of the HTA, and the way it is filled during pattern recognition, lend themselves very naturally to parallel hardware implementations using modern FPGA's with the promise of providing time performances adequate for low level track triggering at future colliders.
    We expect R\&D activities to be spawned in this direction as a follow up to this work.


\acknowledgments

We thank the U.S. Muon Acceleration Program for providing the beam-induced background sample and the International Muon Collider Collaboration for the simulation and reconstruction code.
This work was supported by the European Union’s Europe Research and Innovation programs through the Research Infrastructures INFRADEV Grant Agreement No. 101094300. M.C. received financial support from the European Union’s Horizon 2020 Marie Sklodowska-Curie RISE Grant Agreement No. 101006726.






\end{document}